\documentclass[twocolumn, times, tighten,twocolappendix, a4paper]{aastex701}
\usepackage{amsmath,amssymb}
\usepackage{txfonts}
\usepackage[T1]{fontenc}
\usepackage{graphicx,multirow,hyperref,url,color,xspace}
\DeclareMathAlphabet      {\mathbfit}{OML}{cmm}{b}{it}

\newcommand{\g}{$\gamma$}
\newcommand{\msun}{{M}_{\sun}}
\newcommand{\rsun}{{R}_{\sun}}

\newbox\grsign \setbox\grsign=\hbox{$>$} \newdimen\grdimen \grdimen=\ht\grsign
\newbox\simpropbox
\setbox\simpropbox=\hbox{\raise.5ex\hbox{$\propto$}\llap
 {\lower.5ex\hbox{$\sim$}}}\ht2=\grdimen\dp2=0pt

\begin{document}
\defcitealias{LHAASO26}{L25}
\newcommand{\lh}{{\citetalias{LHAASO26}}\xspace}

\title{The counterjet dominates the production of PeV photons from Cyg X-3}

\author[0000-0002-0333-2452]{Andrzej A. Zdziarski}
\affiliation{Nicolaus Copernicus Astronomical Center, Polish Academy of Sciences, Bartycka 18, PL-00-716 Warszawa, Poland} 
\email[show]{aaz@camk.edu.pl}

\author[0000-0003-0102-5579]{Anton Dmytriiev}
\affiliation{Wits Centre for Astrophysics, School of Physics, University of the Witwatersrand, Private Bag 3, Johannesburg 2050, South Africa} \email[show]{amdmame@gmail.com}

\author[orcid=0000-0002-9677-1533]{Karri I. I. Koljonen}
\affiliation{Institutt for Fysikk, Norwegian University of Science and Technology, H\"{o}gskoleringen 5, Trondheim, 7491  Norway}
\email[show]{karri.koljonen@ntnu.no}  

\begin{abstract}
We study the physical mechanisms underlying the production of orbitally modulated PeV photons from Cyg X-3, recently discovered by the LHAASO collaboration. Our key findings are as follows. Helium nuclei are accelerated in a compact, strongly magnetized region within the jet, but they then advect quickly downstream into regions with a weaker field, allowing them to diffuse out of the jet, where they produce pions in hadronic collisions with both the stellar photons and the stellar wind of the Wolf-Rayet donor. The optical depths across the binary are $\lesssim$1 for both types of interactions, implying that their rates are proportional to the column densities along the particle paths. Given the low viewing angle of Cyg X-3 ($i\approx\!26\degr$--$28\degr$), most of the observed photons are produced by the relativistic hadrons accelerated in the counterjet (for which the column densities toward the observer are much longer than for the jet). This also explains why the phase-folded PeV photon flux peaks on the opposite side of superior conjunction from the (also orbitally modulated) GeV photons, which are produced by collisions of relativistic electrons with stellar photons in the optically thick regime. This then implies that the GeV emission is produced in the approaching jet. 
\end{abstract}

\section{Introduction} \label{Intro}

\begin{figure*}
\centerline{\includegraphics[width=11cm]{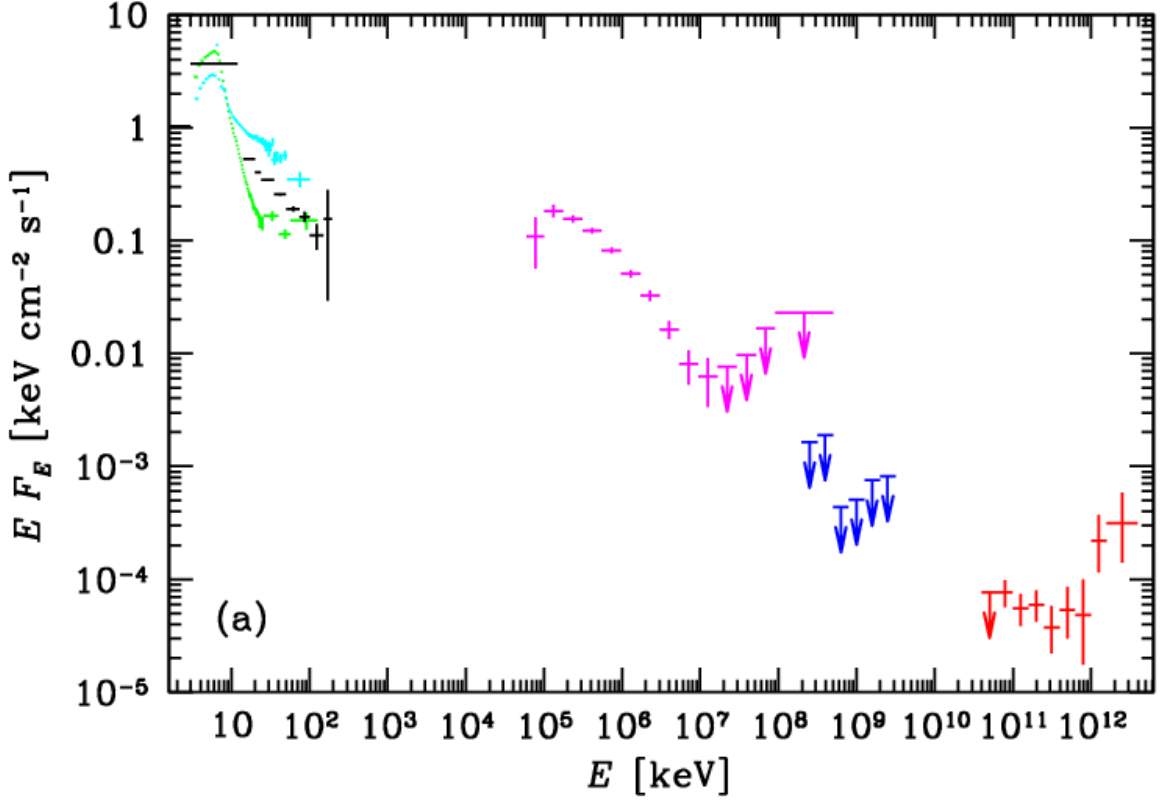}
\includegraphics[width=7.cm]{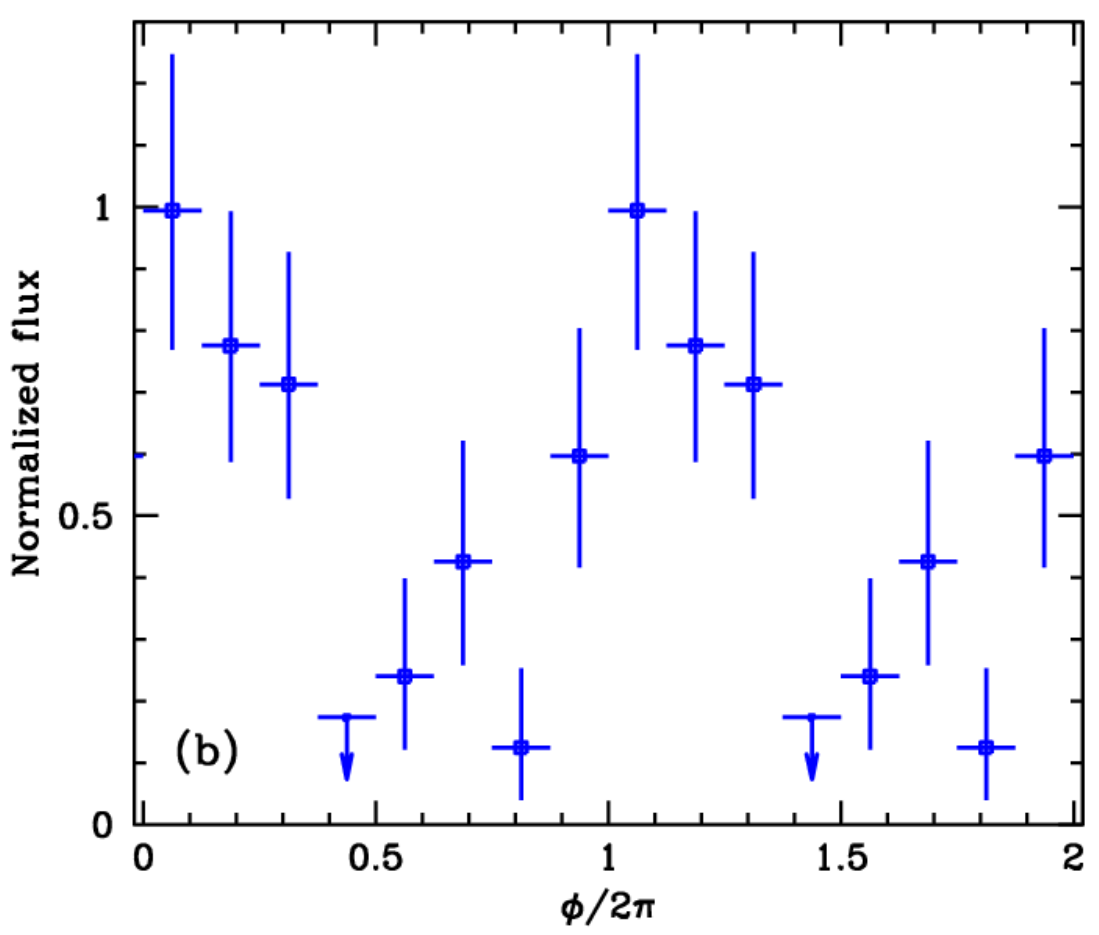}}
\caption{(a) The broad-band spectrum of Cyg X-3 in its $\gamma$-ray flaring state, which corresponds to the soft state in X-rays. The LHAASO PeV data, corrected for CMB absorption (red), are from \lh. We show two representative soft-state spectra in the 3--$10^2$ keV range from \citet{SZM08}, the GeV spectrum (magenta) from \citet{Dmytriiev24}, and the TeV upper limits (blue) from \citet{Aleksic10}. (b) The orbital-phase-folded and normalized light curve of the photon flux from Cyg X-3 in its $\gamma$-ray flaring state for photon energies $E\geq$0.1 PeV, as observed (not corrected for CMB absorption). The normalization factor is $1.05\times 10^{-15}$ cm$^{-2}$ s$^{-1}$. Adapted from \lh. 
}\label{SED_lc}
\end{figure*}

The high-mass binary Cyg X-3 is among the first X-ray binaries discovered \citep{Giacconi67}. Despite many years of intensive research, it remains a puzzling system. It hosts either a neutron star or a black hole, though a black hole appears more likely; see \citet{Koljonen17} and \citet{Antokhin22} for discussions. The donor of Cyg X-3 is a Wolf-Rayet (hereafter WR) star \citep{vanKerkwijk92, vanKerkwijk96, Koljonen17}. This high-mass X-ray binary is unusually compact, with an orbital period of $P_{\rm orb}\approx 4.8$ h (\citealt{Antokhin19} and references therein). This implies a semi-major axis of $a\approx 2.7\times 10^{11}(M/20\msun)^{1/3}$ cm, where $M$ is the total mass. Cyg X-3 appears to be seen at a low orbital inclination; \citet{Antokhin22} estimated $i=29.5\degr\pm 1.2\degr$ by modeling the average infrared (IR) and X-ray light curves, while \citet{Veledina24b} constrained the viewing angle to $26\degr\leq i\leq 28\degr$ to explain the measured very high linear X-ray polarization degree ($>20\%$; \citealt{Veledina24}). The distance to Cyg X-3 was determined from radio parallax as $D\approx 9.7\pm 0.5$ kpc and from Galactic proper motion and line-of-sight radial velocity measurements as $9 \pm 1$ kpc \citep{Reid23}. We assume $D=9$\,kpc and $i=28\degr$ henceforth. 

A major recent discovery for this system was the detection of ultra-high-energy $\gamma$-rays in the $0.06\lesssim E\lesssim 4$ PeV (\citealt{LHAASO25}, hereafter \lh). (Hereafter, $E$ denotes the photon energy in the observer's frame.) The emission occurred only during GeV-photon flares observed simultaneously by the Fermi satellite. As with the GeV emission \citep{Dubus10, Zdziarski12a, Zdziarski18, Dmytriiev24}, the PeV emission was modulated at the system's orbital period (at 3.2$\sigma$ significance; \lh). A compilation of the broadband spectra in the $\gamma$-ray flaring state (from different flaring episodes) is shown in Figure \ref{SED_lc}(a), and the PeV orbital modulation is shown in Figure \ref{SED_lc}(b). The PeV emission has been attributed by \lh to photopion production on both the stellar UV and disk X-ray photons, $p+\gamma\rightarrow p+ \pi^0$, followed by the decay of $\pi^0$ into two $\gamma$-rays. The orbital modulation at $E\gtrsim$0.4 PeV was attributed to the angle between the incoming stellar photons and the line of sight to the observer, which varies with orbital phase. 

Figure \ref{SED_lc}(b) shows that the modulation appears relatively deep. While \lh found that all photons with energies in the range 0.4--4 PeV were detected during the orbital phase range of 0.8--1.3, we have calculated, using the observed spectrum (\lh), that almost 80\% of the photons included in the folded light curve have energies $<$0.4 PeV. Thus, the apparently high depth of the total modulation suggests that those photons were also orbitally modulated. In that case, invoking interactions with X-ray photons from an accretion disk in the model (\lh) may pose a problem because the X-ray emission is axially symmetric about the orbital axis. 

Studies of Cyg X-3 focusing on the physics of hadron acceleration and photopion production were conducted by \citet{Kachelriess25a, Kachelriess25b}. Here, we identify the key physical processes underlying PeV photon production. We also specify the physical conditions in the binary, particularly the structures of the photon field and the WR star's wind. We calculate the orbital modulation of PeV photons and the power required to accelerate hadrons, both of which are needed to account for the observed fluxes. Except for an illustrative example, we do not consider the effects of jet misalignment or bulk motion, since detailed fitting of the observed spectrum and orbital profile is beyond the scope of this Letter. Such an analysis will be presented in the following paper.

\begin{figure*}
\centerline{\includegraphics[width=11cm]{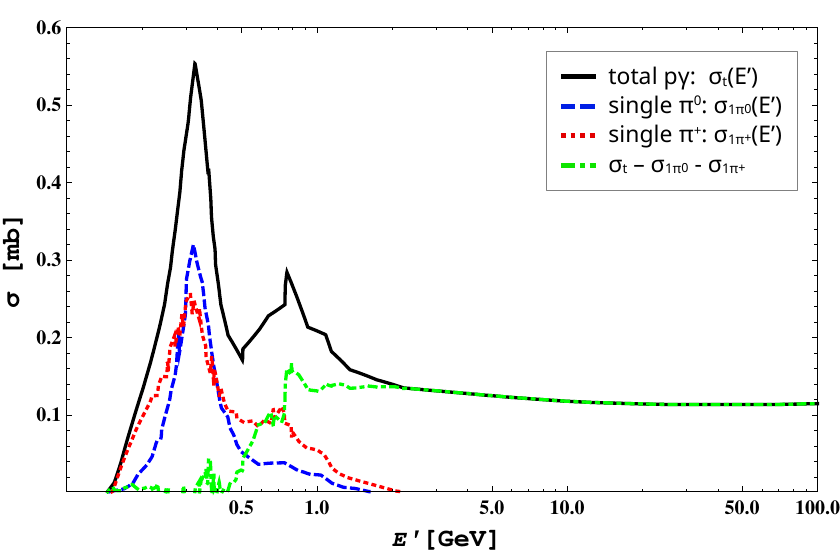}}
\caption{The $p$-$\gamma$ cross sections in mb as functions of the $\gamma$-ray energy in GeV in the hadron rest frame, $E'$. The curve for all inelastic $p$-$\gamma$ interactions is shown in solid black. The curves for single $\pi^0$ and $\pi^+$ production are shown in dashed blue and dotted red, respectively. The green dot-dashed curve shows the difference between the total cross section and the cross sections for single neutral and charged pion production (which includes a minor contribution from diffractive scattering; see figure 3 of \citealt{Mucke00}). Adapted from \citet{Kelner08} and \citet{Mucke00}. 
}\label{p_gamma}
\end{figure*}

\section{The physical processes}
\label{physical}

\subsection{Hadron-induced interactions}
\label{hadron}

The total cross section for photo-pion production in the proton's rest frame, $\sigma_{\rm t}(E')$, including single and multiple $\pi^0$ and $\pi^\pm$ production \citep{Mucke00}, is shown graphically by \citet{Kelner08}, and we show it as the black line in Figure \ref{p_gamma}. We also show the cross sections \citep{Mucke00} for single $\pi^0$ and $\pi^+$ production, $\sigma_{1\pi^0}$ and $\sigma_{1\pi^+}$, as the blue dashed and red dotted curves, respectively. The thresholds for single $\pi$ production in the proton rest frame are $E'_{\rm th}\approx 0.145$ GeV and $\approx$0.150 GeV for neutral and charged pions, respectively. 

 As shown in Figure \ref{p_gamma}, the cross-section for single $\pi^0$ production is approximately half of the total at the peak, but it drops to nearly zero at high energies. The cross-section for single $\pi^+$ production, $p+\gamma \rightarrow n + \pi^+$, is similar, and multiple pion production, with $\sigma_{\rm m}$, dominates at high energies \citep{Mucke00}. We assume that all of the events producing multiple pions include the production of at least one $\pi^0$. Then, the combined cross-section for producing either a single $\pi^0$ or a $\pi^0$ accompanied by charged pions is
 \begin{equation}
     \sigma_{\pi^0}\approx \sigma_{1\pi^0}+\sigma_{\rm m}.
     \label{sigmapi0}
 \end{equation}
The fraction of proton energy going into pion production, i.e., the inelasticity, increases from $K_\pi \approx 0.2$ in the $\Delta$-resonance regime to $K_\pi \sim 0.5$ in the multi-pion regime \citep{Kelner08}. Given the smooth transition between these regimes as a function of the logarithm of energy, we can adopt a phenomenological interpolation between the two limits.
\begin{equation}
K_\pi(E')\approx\begin{cases}
0.2, &0.145\lesssim E'\lesssim 0.5\,{\rm GeV};\\
0.2+0.13\ln\frac{E'}{0.5\,{\rm GeV}}, &0.5\lesssim E'\lesssim 5\,{\rm GeV};\\
0.5, &E'\gtrsim 5\,{\rm GeV}.
\label{inelasticity}
\end{cases}
\end{equation}
Up to $\approx$0.5 GeV, individual pions carry away about $\approx$20\% of the initial proton energy, and this fraction decreases to $\approx$10\% at high energies. A neutral pion decay, $\pi^0\rightarrow \gamma +\gamma$, produces two photons, each carrying away on average $\approx$10--5\% of the proton energy. 
 
 Charged pions decay into muons and neutrinos, $\pi^+\rightarrow \mu^+ +\nu_{\mu}$, with the muon carrying $\approx$80\% of the pion energy. The muon then decays into a positron and two neutrinos, $\mu^+\rightarrow e^+ +\nu_{e}+\bar{\nu_\mu}$, with each carrying about 1/3 of the muon energy. Thus, a positron receives $E_e\approx 5$--2.5\% of the proton energy, i.e., about half the average energy of photons from the $\pi^0$ decay. 

 The positrons then both lose energy by emitting synchrotron radiation and by Compton up-scattering soft photons present in the source. The fractional electron synchrotron energy loss per unit distance is
\begin{equation}
\frac{\dot \gamma_e}{c\gamma_e}=\frac{\sigma_{\rm T}\gamma_e B^2}{6\pi m_{\rm e}c^2}\approx 8.5\times 10^{-7}{\rm cm}^{-1} \frac{E_e}{10^{15}\,{\rm eV} }
\left(\frac{B}{100\,{\rm G}}\right)^2,
\end{equation}
where $B$ is the magnetic field strength in the emission region (where the Hillas criterion does not apply; see Section \ref{regions} below). An order-of-magnitude estimate of the magnetic field strength within the binary and outside both the jet and the donor, based on a model of field advection that explained the long-term time lags observed in Cyg X-3, was $B\sim 100$ G \citep{CZ20}. Because the system's characteristic dimension is the semi-major axis, electron synchrotron losses are very significant unless the magnetic field is very weak. 

On the other hand, there is virtually no constraint on the magnetic field strength from the synchrotron emission of either protons or He nuclei. That emission is much weaker than that of the electrons, and the resulting limit is $B\lesssim 3\times 10^9$ G (modifying the result of \citealt{DeJager96} for He nuclei at 20 PeV). 

There will also be some scattering of IR photons (see Section \ref{photon}). The scattering occurs in the extreme Klein-Nishina regime, where the cross section is \citep{RL79},
\begin{equation}
 \sigma_{\rm KN}\approx\frac{3\sigma_{\rm T} m_{\rm e}c^2}{8 E'}\left(\ln \frac{2 E'}{m_{\rm e}c^2}+\frac{1}{2}\right),
 \label{skn}
\end{equation}
where $E'=\gamma_e E (1-\beta_e\cos\alpha)$ is the photon energy in the electron rest frame, $m_{\rm e}$, $\gamma_e\equiv E_e/m_{\rm e} c^2$, and $\beta_e$ are the electron's rest mass, Lorentz factor, and velocity in units of $c$, respectively; $\alpha$ is the scattering angle (in the LAB frame); and $\sigma_{\rm T}$ is the Thomson cross section. Since $\sigma_{\rm T}=665$ mb, the Klein-Nishina cross section in our case is comparable to, or even larger than, that for photopion production. Scattering in the Klein-Nishina regime yields photons with energies only moderately lower, by a factor of $\gtrsim$0.6 for $\gamma_e E/m_{\rm e}c^2>10$, than the electrons' energies \citep{BG70}. However, at $B\sim 100$ G, synchrotron energy loss will be four orders of magnitude faster, and we neglect this channel of $\gamma$-ray production hereafter.

Importantly, the donor of Cyg X-3 is a WR star; thus, the main species in the accretion flow and jet is He. Consequently, most accelerated nuclei are He rather than protons. However, the binding energy of He is 28.3 MeV, substantially below the threshold for pion production. In addition, the cross-section for photodisintegration in the giant dipole resonance region is $\sim$1 mb, larger than that for photo-pion production, and it peaks around 25--30 MeV in the nucleus rest frame before gradually decreasing at higher energies (\citealt{Raut12} and references therein). Therefore, most accelerated He nuclei will photodisintegrate before producing pions. Importantly, full photodisintegration of each He nucleus yields two neutrons, while partial disintegration channels typically produce at least one neutron. In addition, neutrons are produced in photo-pion interactions of charged hadrons through the $\pi^+$ channel. Since neutrons are electrically neutral, they are not confined by the magnetic field of the acceleration region and can escape freely. Their decay time in the LAB frame is $\gamma_n 878$ s (where $\gamma_n$ is the neutron Lorentz factor), which is much longer than the dynamical time-scale of the system. These neutrons subsequently interact with the ambient radiation field and produce neutral ($n+\gamma\rightarrow n+ \pi^0$) and charged ($n+\gamma \rightarrow p + \pi^-$) pions, with cross-sections comparable in magnitude to those of the $p+\gamma$ reaction \citep[e.g.][]{Briscoe2019}. Then, $\pi^-$ decay similarly to proton-produced pions, except that they produce an electron instead of a positron.

The donor star in Cyg X-3 also shows a strong wind; see Section \ref{photon} below. Hadrons accelerated to PeV energies will produce pions in collisions with He nuclei in the wind. The inelastic cross-section for proton-proton interactions in the 1--20 PeV energy range is $\sigma_{\rm pp} \sim 60$--75 mb \citep[e.g.][]{Kelner2006, Kafexhiu2014}, and increases only weakly with energy. For proton-nucleus interactions, the cross-section can be estimated geometrically assuming a constant nuclear density, so the nuclear radius scales as $r_{\rm A} \propto A^{1/3}$. This yields $\sigma_{\rm pA} \propto r_{\rm A}^2 \approx A^{2/3} \sigma_{\rm pp}$, resulting in $\sigma_{\rm pHe} \sim 150$--200 mb. The threshold for pion production in this process is $\approx$280 MeV in the lab frame, well below the range of the hadron energies we consider, implying that all of the accelerated hadrons considered here will be able to produce pions. In particular, protons with energies of ${\cal E}_{\rm p}=1$--20 PeV produce pions with a multiplicity of $\approx$50--100 \citep[e.g.][]{Kafexhiu2014}. Approximately equal fractions of $\pi^0$, $\pi^+$, and $\pi^-$ are produced, so about 1/3 of the pions will be neutral. Thus, all of these interactions will produce some $\pi^0$ pions. The inelasticity of this process is large, $\approx 0.5$--0.6. The most energetic pions carry up to $\sim 10$\% of the primary proton energy, ${\cal E}_\pi\approx 0.1 {\cal E}_{\rm p}$, resulting in $\gamma$-rays from $\pi^0$ decay with energies up to $\sim 0.05 {\cal E}_{\rm p}$, which can cover the range of $E>0.06$ PeV (corresponding to the LHAASO detected photons). 

\subsection{The photon and wind emission of the donor}
\label{photon}

\begin{figure}
\centerline{\includegraphics[width=7.5cm]{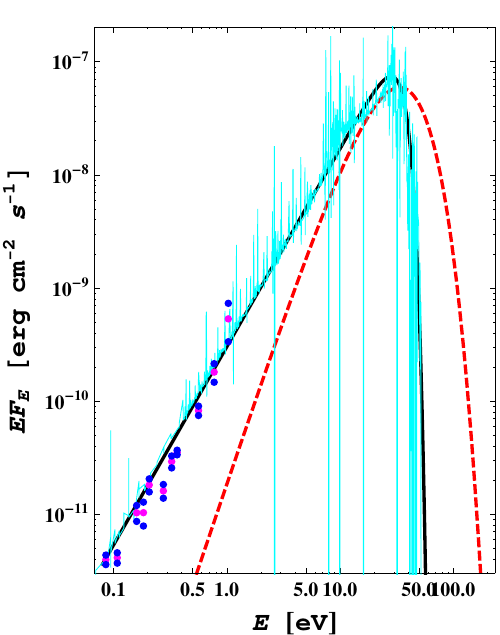}}
\caption{The IR-to-UV spectrum of Cyg X-3 as seen at 9 kpc. The cyan symbols show the adopted model spectrum with the hydrostatic core at $R_*=1\times 10^{11}$\,cm and $T_*=1\times 10^5$\,K (see Section \ref{photon}). This spectrum is significantly softer than the blackbody at these $R_*$ and $T_*$, shown by the red dashed curve. The IR measurements from UKIRT and ISO are shown as magenta dots, with blue dots indicating the measurement errors. The black solid curve shows our phenomenological approximation to the spectrum, given by Equation (\ref{efe}). 
}\label{IRUV}
\end{figure}

The next ingredient for the production of $\gamma$-rays is the photon field within the binary. For the WR emission, we assume the parameters of the hydrostatic core, with radius $R_*=1.5\rsun \approx 1.04\times 10^{11}$\,cm and effective temperature (defined by the observed stellar luminosity $L=4\pi \sigma_{\rm SB} R_*^2 T_*^4$, where $\sigma_{\rm SB}$ is the Stefan-Boltzmann constant) of $T_*=1\times 10^5$\,K \citep{Koljonen17}. This model\footnote{We used the WNL-H20E model 13-19 from \url{https://www.astro.physik.uni-potsdam.de/PoWR}\label{powr}} also assumes $M=12\msun$, and the resulting mass loss rate is $\dot M_{\rm w}= 10^{-4.74}\msun$/yr. However, the core's emission passes through the dense stellar wind, which softens it. The wind is optically thick up to a large distance, with the radius at which the optical depth is 2/3 of $R_{2/3}\approx 5.2\times 10^{11}$ cm, which is $\approx 1.9 a$, i.e., substantially more than the distance of the compact object from the donor's center. The resulting model spectrum \citep{Hamann04, Todt15}, as seen at a distance of 9 kpc, is shown in Figure \ref{IRUV} as cyan symbols. Its effective temperature is $T_{{2/3}} \approx 4.5\times 10^4$ K, but its spectral shape differs significantly from that of a blackbody; we see both strong IR emission and a sharp high-energy cutoff. The excess IR emission with respect to the blackbody is mostly due to bremsstrahlung from the accelerated wind. 

The IR emission was also observed. We use the Cyg X-3 measurements in the quiescent state obtained with the United Kingdom Infrared Telescope (UKIRT; \citealt{Fender96}) and the Infrared Space Observatory (ISO; \citealt{Ogley01}) in the 0.085--1.03 eV energy range, shown as magenta dots in Figure \ref{IRUV}. The blue dots indicate the measurement errors. The IR emission has been dereddened\footnote{We note that \citet{Ogley01} analyzed the same data using the extinction law of \citet{Mathis90}, which has been shown to be incorrect by \citet{Stead09}. } using the extinction law of \citet{Stead09} with $A_K=1.4\pm 0.1$ \citep{Koljonen17}. The IR measurements agree well with the model spectrum. We can approximate the model spectrum by
\begin{equation}
    E F(E)\approx F_0 \left(\frac{E}{1\,{\rm eV}}\right)^g \exp\left[-\left(\frac{E}{E_{\rm c}}\right)^b\right],
    \label{efe}
\end{equation}
where $F_0=3.0\times 10^{-10}$ erg cm$^{-2}$ s$^{-1}$, $g=1.75$, $E_{\rm c}=30$ eV, and $b=5$. It is shown in Figure \ref{IRUV} as the black solid curve. The corresponding differential photon density at $r>R_{2/3}$ follows from integrating the blackbody specific intensity over the solid angle subtended by the star, analogous to the case of the flux from a uniform sphere (see \citealt{RL79}). This yields 
\begin{equation}
n(E,R)=\frac{2 F(E)}{E c} \left(D\over R\right)^2\left[1-\sqrt{1-\left(R_*\over R\right)^2}\right], \quad R>R_{2/3}.
\label{ene}
\end{equation}

\begin{figure}
\centerline{\includegraphics[width=7.5cm]{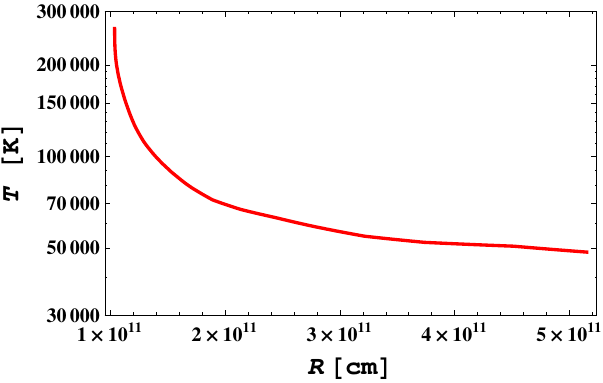}}
\caption{The radial dependence of the wind temperature between $R_*$ and $R_{2/3}$ for the model used here, see footnote \ref{powr}.
}\label{temperature}
\end{figure}

On the other hand, the photon density inside $R_{2/3}$ increases as the radius decreases and as the optical depth, $\tau$, measured from $R_{2/3}$ downward, increases. The local effective temperature at radius $R$ follows from the constant energy flux, $L=4\pi \sigma_{\rm SB} R^2 T_{\rm eff}^4$. However, the actual local temperature is higher due to radiative diffusion. Here, we use the electron temperature as a function of radius, $T(R)$, tabulated on the web page given in footnote \ref{powr}, and we show it in Figure \ref{temperature}. The photon distribution is approximately given by the blackbody law at the local temperature, 
\begin{equation}
E n(E,R)\approx \frac{8\pi}{(h_{\rm P}c)^3} \frac{E^3}{\exp[E /k T(R)]-1}, 
\label{bb}
\end{equation}
where $h_{\rm P}$ is the Planck constant. Departures from this law are relatively minor, as we verified using figure 6 of \citet{Sander20}. 

The electron and He number densities of a line-driven wind (averaged over clumping and assuming spherical symmetry) at $R$ are related to the wind velocity, $v(R)$, and the stellar mass-loss rate, $\dot M_{\rm w}$, via the continuity equation, 
\begin{equation}
n_{\rm e}(R) = {\dot M_{\rm w} \over 4 \pi m_{\rm p}\mu_{\rm e} R^2 v(R) },\quad n_{\rm He}(R)=\frac{1}{2} n_{\rm e}(R),
\label{n_r}
\end{equation}
respectively, where $\mu_{\rm e}\approx 2$ is the mean electron molecular weight. We use the standard parametrization for the wind velocity, 
\begin{equation}
v(R) = v_{\infty} (1 - R_{\star}/R)^{\beta_{\rm w}},\quad R>R_*,
\label{velocitylaw}
\end{equation}
where $v_{\infty}$ is the wind terminal velocity and $\beta_{\rm w}\approx 1$--2. The model we use has $\beta_{\rm w}=1$ \citep{Todt15} and $v_{\infty}=1.0\times 10^8$ cm s$^{-1}$.

Accelerated hadrons will produce pions in collisions with He nuclei in the wind (see Section \ref{hadron}). Given the strength of the wind in Cyg X-3, the optical depth for this process can be comparable to unity (see Section \ref{depths}), indicating that it is an important process. Because its cross-section is approximately independent of energy, it will produce pions over a broad range of energies. Its rate will be modulated by the orbit because the optical depth through the wind toward the observer depends on the orbital phase.

\section{The acceleration and emission regions}
\label{regions}

The site of the hadron acceleration in Cyg X-3 is located most likely within the jet. The jet moves with a relativistic velocity, implying the hadrons are advected away from the compact object. The crucial issue is whether the hadrons are in the slow- or fast-cooling regime. In the latter case, the hadrons would lose their energy within the acceleration site, and thus the acceleration and emission regions would be nearly coincident. In the former case, the hadrons will be advected downstream in the jet and emit outside the acceleration site. Below, we find that the hadrons are in the slow-cooling regime. If the jet is conical, the magnetic field quickly declines downstream, with the toroidal and poloidal field strengths of $B\propto h^{-1}$ and $\propto h^{-2}$, respectively.

The size of the {\it acceleration\/} region, $R_{\rm acc}$, is constrained by the requirement that particles remain magnetically confined during acceleration. In its general form, the Hillas criterion includes a factor $\beta_{\rm sc}$ representing the characteristic velocity of the scattering centers (or, equivalently, the effective accelerating electric field), such that ${\cal E}_{\rm max} \approx Z e B R_{\rm acc} \beta_{\rm sc}$ \citep{Hillas84}. In the context of diffusive shock acceleration, we adopt $\beta_{\rm sc} \sim \beta_{\rm sh}$, where $\beta_{\rm sh} c$ is the shock velocity. The corresponding acceleration region size is then
\begin{equation}
    R_{\rm acc}\equiv \frac{{\cal E}_{\rm max}}{Z e B \beta_{\rm sh}}\approx 5.5\times 10^{10}\, {\rm cm}\,\frac{{\cal E}_{\rm max}}{20\,{\rm PeV}}\frac{2}{Z} \frac{10^3\,{\rm G}}{B} \frac{0.6}{\beta_{\rm sh}},
    \label{rlarmor}
\end{equation}
where $Z$ is the particle's atomic number. More generally, efficient acceleration requires that the acceleration timescale be shorter than both the escape time from the acceleration region and the advection time along the jet. For diffusive shock acceleration (first-order Fermi), the acceleration timescale can be written as $t_{\rm acc} \approx \eta {\cal D}/(c^2\beta_{\rm sh}^2)$ (e.g., \citealt{Drury1983, WeidingerSpanier2015}). Here, ${\cal D}$ is the diffusion coefficient, and $\eta$ is a factor of a few that accounts for shock compression and geometry. In the Bohm limit, ${\cal D} \approx (1/3) R_{\rm L} c$, with $R_{\rm L}$ denoting the particle's Larmor radius, yielding $t_{\rm acc} \approx \eta R_{\rm L} / (3\beta_{\rm sh}^2 c)$. The escape time depends on the confinement regime. In the diffusive limit ($R_{\rm L} \ll R_{\rm acc}$), $t_{\rm esc} \approx R_{\rm acc}^2/{\cal D}$, while in the ballistic limit ($R_{\rm L} \sim R_{\rm acc}$), applicable to particle escape near ${\cal E}_{\rm max}$, one expects $t_{\rm esc} \sim R_{\rm acc}/c$. The advection time along the jet is $t_{\rm adv}\sim \Delta h/\beta_{\rm j}c$, where $\beta_{\rm j}$ is the jet velocity and $\Delta h$ is the length of the acceleration zone along the jet. Requiring $t_{\rm acc} < t_{\rm esc}$ in the ballistic regime implies a lower limit on the magnetic field strength,
\begin{equation}
B\gtrsim \frac{\eta}{3} \frac{{\cal E}_{\rm max}}{\beta_{\rm sh}^2 Z e R_{\rm acc}},
\label{Bmin}
\end{equation}
where we assumed $\Delta h/\beta_{\rm j}c>R_{\rm acc}/c$. With $R_{\rm acc}=10^{10}$ cm, $Z=2$, ${\cal E}_{\rm max}=20$ PeV, $\eta/3 \sim 1$, and $\beta_{\rm sh} \sim 0.6$ (comparable to the jet velocity $\beta_j$; \citealt{Dmytriiev24}), we obtain $B\gtrsim 9\times 10^{3}$ G. 

We compare it to the limiting field for pair production by $\gamma$-rays in a magnetic field \citep{Erber66}, $\gamma+ B\rightarrow e^+ + e^- + B$. An approximate threshold for this reaction is $(B_\perp/B_{\rm cr})(E/m_{\rm e}c^2)\gtrsim 0.1$, where $B_\perp$ is the magnetic field strength perpendicular to the photon momentum, $B_{\rm cr}=2\pi m_{\rm e}^2 c^3/e h_{\rm P}\approx 4.14\times 10^{13}$ G is the critical magnetic field, and $e$ is the electron charge. The reported detection of $E_{\rm max}\approx 4$ PeV photons (\lh) may appear to imply $B_\perp\lesssim 10^3$ G, in conflict with the Hillas limit. However, only a tiny fraction of the accelerated hadrons will emit $\gamma$-rays within the acceleration site. Most of the accelerated hadrons are advected downstream to regions with weaker magnetic fields, where they do not need to be confined. Thus, this constraint is of no practical consequence for Cyg X-3.

We also note that the Hillas limit yields an upper bound on the magnetic jet power in the acceleration region, independent of both $B$ and $R_{\rm acc}$,
\begin{align}
    P_{B}&=2\pi R_{\rm acc}^2 c \frac{B^2}{4\pi} \beta \Gamma^2\gtrsim \frac{{\cal E}_{\rm max}^2 c\beta_{\rm j} \Gamma_{\rm j}^2}{2 Z^2 e^2}\label{PB}\\ &\approx1.7\times 10^{37}\left(\frac{{\cal E}_{\rm max}}{20\,{\rm PeV}}\right)^2 \left(\frac{2}{Z}\right)^2 \left(\frac{\eta}{3}\right)^2 \beta_{\rm sh}^{-4} \beta_{\rm j}\Gamma_{\rm j}^2\,\frac{{\rm erg}}{{\rm cm}^2{\rm s}},\nonumber
\end{align}
where $\Gamma_{\rm j}$ is the jet bulk Lorentz factor, and the toroidal magnetic field is assumed. This limit is moderate, given that the accretion rate in Cyg X-3 has been estimated to be super-Eddington \citep{Veledina24}. The jet bulk motion in Cyg X-3 appears moderate, with $\beta_{\rm j}\Gamma_{\rm j}^2\sim 1$ (e.g., \citet{Dmytriiev24}).

\section{The model}\label{model}

\subsection{Geometry and transformations}
\label{geo}

\begin{figure*}
\centerline{\includegraphics[width=11cm]{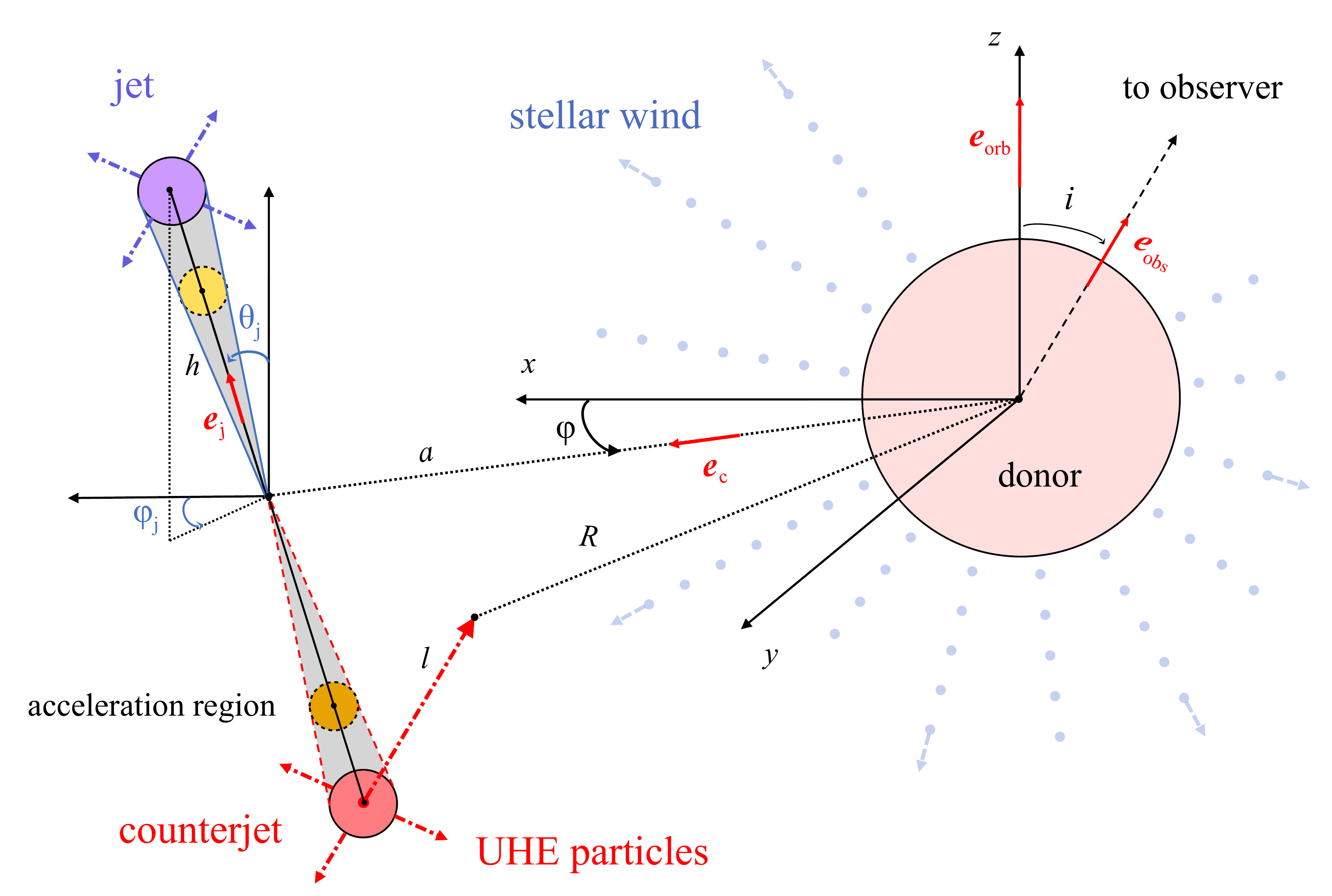}}
\caption{The geometry of the binary and the jets. The $x$ and $y$ axes lie in the binary plane, and $+z$ aligns with the binary vector. The observer is at an angle $i$ with respect to ${\mathbfit e}_{\rm orb}$, and $\phi$ is the orbital phase; $\phi=0$ and $\pi$ correspond to the superior and inferior conjunctions, respectively. The binary rotation follows increasing $\phi$ and is assumed to be counterclockwise. Then, $\theta_{\rm j}$ is the inclination of the jet vector, ${\mathbfit e}_{\rm j}$, with respect to ${\mathbfit e}_{\rm orb}$, $\phi_{\rm j}$ is its azimuthal angle, and $h_{\rm j}$ is the distance from the hadron emission point to the center of the compact object. The distance from the emission region to a point along the line to the observer is $l$. The distance from that point to the donor's center is $R$. The acceleration region is upstream of the emission region.
} \label{geometry}
\end{figure*}

The jet is inclined relative to the binary axis, with azimuthal and polar angles $\theta_{\rm j}$ and $\phi_{\rm j}$, respectively. The binary rotation is counterclockwise on the sky \citep{Veledina24}, and $\phi$ is the phase measured from superior conjunction. The geometry is shown in Figure \ref{geometry}. 

The protons, He nuclei, and neutrons are ultrarelativistic, and the kinematics imply that the directions of the produced pions, $\gamma$-rays, and electrons are nearly aligned with the hadron direction. We consider hadrons with Lorentz factors $\gamma_{\rm h} \gg 1$ leaving the jet at height $h$ along the jet. We then assume that both neutrons and charged ions propagate in straight lines, neglecting the curvature of the latter's paths along the magnetic field lines. However, $B\sim 100$ G will cause relatively modest path curvature of $R_{\rm L}\sim a$ at ${\cal E}=20$ PeV. We also consider the Lorentz factors in the LAB frame rather than in the jet frame. Thus, these factors include both the jet motion and the hadron's motion relative to the jet. 

We define unit vectors pointing towards the observer, from the stellar center to the compact object, and along the jet,
\begin{align}
&{\mathbfit e}_{\rm obs}=(-\sin i, 0, \cos i),\quad {\mathbfit e}_{\rm c}=(\cos\phi,\sin\phi, 0),\nonumber\\
&{\mathbfit e}_{\rm j}=(\sin\theta_{\rm j}\cos\phi_{\rm j}, \sin\theta_{\rm j}\sin\phi_{\rm j} , \cos\theta_{\rm j}),\label{vectors}
\end{align}
respectively. Then, the vector connecting the donor center to a point along the hadron trajectory at a distance $l$ from the emission point, and the corresponding unit vector, are
\begin{equation}
{\mathbfit R}=
a {\mathbfit e}_{\rm c}+ h {\mathbfit e}_{\rm j}+ l {\mathbfit e}_{\rm obs},\quad {\mathbfit e}_{\rm R}={\mathbfit R}/|{\mathbfit R}|,
\label{interaction}
\end{equation}
respectively. The length of ${\mathbfit R}$ is given by,
\begin{align}
&R^2 = (-l \sin i+a \cos\phi+h \sin\theta_{\rm j}\cos\phi_{\rm j})^2
\nonumber \\
&+(h \sin\theta_{\rm j}\sin\phi_{\rm j}+a \sin\phi)^2+(l \cos i +h \cos\theta_{\rm j})^2. \label{r2}
\end{align} 

To simplify treatment of the scattering angles, we approximate the emission at the interaction point as originating from a point source at the donor's center. Therefore, for a given $l$, there is a single angle $\alpha$ between the directions of the hadron and the photon, given by
\begin{align}
&\cos \alpha={\mathbfit e}_{\rm R}\cdot {\mathbfit e}_{\rm obs}=
\label{cosine}\\
&\frac{\cos i(l \cos i+h \cos\theta{\rm j})+\sin i(l\sin i -h\cos \phi_{\rm j}\sin\theta_{\rm j}-a\cos\phi)}{R}.\nonumber
\end{align}
The photon energy in the hadron rest frame and the threshold energy in the LAB frame are
\begin{equation}
E' =\gamma_{\rm h} E (1-\beta_{\rm h}\cos\alpha),\quad E_{\rm th}=\frac{E'_{\rm th}}{\gamma_{\rm h}(1-\beta_{\rm h} \cos\alpha)},
\label{E_prf}
\end{equation}
respectively, and $\beta_{\rm h}\approx 1$ at $\gamma_{\rm h}\gg 1$. 

\subsection{Optical depths}
\label{depths}

\begin{figure*}
\centerline{\includegraphics[width=\columnwidth]{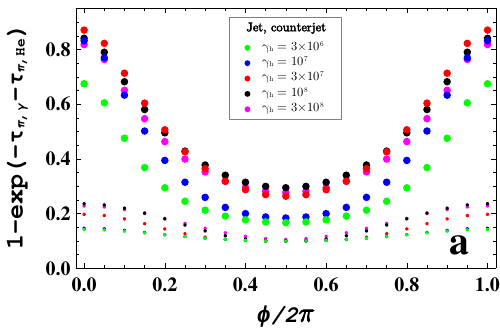}
\includegraphics[width=\columnwidth]{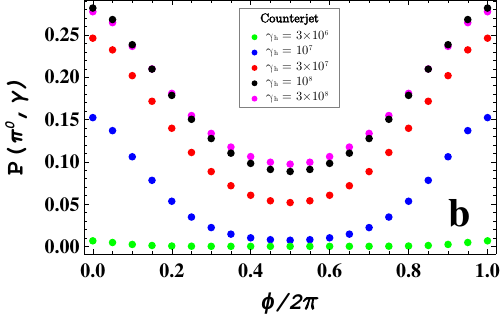}}
\centerline{\includegraphics[width=\columnwidth]{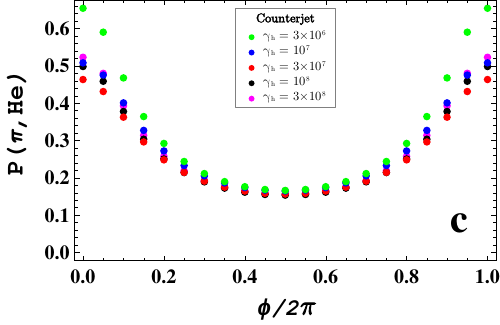}
\includegraphics[width=\columnwidth]{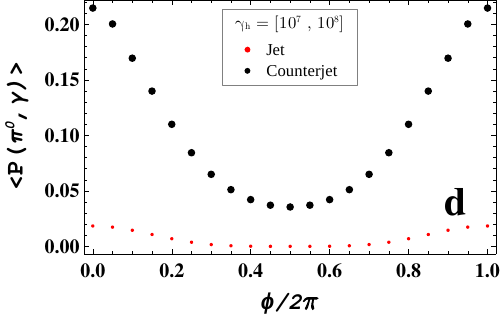}}
\caption{The Interaction probabilities as functions of the orbital phase for (a) the sum of pion production on stellar photons and He nuclei, Equation (\ref{prob}), for the jet (small circles) and the counterjet (large circles). (b) The probability for the production of $\pi^0$ on stellar photons for the counterjet, Equation (\ref{probpi0}). (c) The probability for the production of pions on the He nuclei of the stellar wind, Equation (\ref{probHe}), for the counterjet. The proton Lorentz factors in (a--c) are $\gamma_{\rm h}=3\times 10^6$ (green), $\gamma=10^7$ (blue), $3\times 10^7$ (red), $10^8$ (black), and $3\times 10^8$ (magenta). (d) The probability of producing $\pi^0$ from stellar photons averaged over a power-law distribution, Equation (\ref{average}). The proton Lorentz factors are in the range $10^7$--$10^8$, and the power-law index is $p=2.2$. The small red and large black circles denote the jet and counterjet, respectively. The parameters are $h_{\rm j}=0.7a$, $i=28\degr$, and $\theta_{\rm j}=0$ (for which all of the probabilities are symmetric with respect to $\phi/2\pi=0.5$). 
}\label{tau_plot}
\end{figure*}

The optical depth for any proton interaction outside the acceleration region along $l$ (i.e., assuming propagation along a straight line) with a stellar photon is given by
\begin{equation}
\tau_{\pi,\gamma}(\gamma_{\rm h})=\int_0^\infty\!\! {\rm d}l [1-\beta_{\rm h} \cos\alpha(l)]\! \int_{E_{\rm th}}^\infty \!\!{\rm d}E\, n(E,l)\sigma_{\rm t}[E'(\gamma_{\rm h},E)],
\label{taug}
\end{equation}
and the corresponding one for producing a $\pi^0$ pion is
\begin{equation}
\tau_{\pi^0,\gamma}(\gamma_{\rm h})=\int_0^\infty\!\! {\rm d}l [1-\beta_{\rm h} \cos\alpha(l)]\! \int_{E_{\rm th}}^\infty \!\!{\rm d}E\, n(E,l)\sigma_{\pi^0}[E'(\gamma_{\rm h},E)],
\label{taugpi0}
\end{equation}
The optical depth for proton-He nuclear collisions in the stellar wind (see Section \ref{photon}) is given by
\begin{equation}
\tau_{\pi,{\rm He}} (\gamma_{\rm h})=\int_0^\infty\!\! {\rm d}l \, n_{\rm He}(l)\sigma_{\rm pHe}(\gamma_{\rm h}).
\label{taupHe}
\end{equation}
In our approximation, $\sigma_{\rm pHe}\approx 200$ mb (assumed energy-independent here, given the weak energy dependence of the cross-section at PeV energies), $\tau_{\pi,{\rm He}}$ is independent of $\gamma_{\rm h}$, and the produced pions, and subsequently the $\gamma$-rays, will have a wide range of energies. 

The total interaction probability is
\begin{equation}
    {\cal P}_{\rm total}(\gamma_{\rm h})=1-\exp(-\tau_{\pi,\gamma}-\tau_{\pi,{\rm He}}).
    \label{prob}
\end{equation}
This probability equals the sum over the pion-producing channels. The probability of producing a $\pi^0$ from stellar photons is given by
\begin{equation}
    {\cal P}_{\pi^0,\gamma}(\gamma_{\rm h})=\left[1-\exp(-\tau_{\pi,\gamma}-\tau_{\pi,{\rm He}})\right]\frac{\tau_{\pi^0,\gamma}}{\tau_{\pi,\gamma}+\tau_{\pi,{\rm He}}},
    \label{probpi0}
\end{equation}
and the probability of a hadronic interaction with the wind is
\begin{equation}
    {\cal P}_{\pi,{\rm He}}(\gamma_{\rm h})=\left[1-\exp(-\tau_{\pi,\gamma}-\tau_{\pi,{\rm He}})\right]\frac{\tau_{\pi,{\rm He}}}{\tau_{\pi,\gamma}+\tau_{\pi,{\rm He}}}.
    \label{probHe}
\end{equation}
Note that these probabilities are for the first interaction of a proton. 

Some examples of the resulting orbital modulation profiles are shown in Figure \ref{tau_plot}. We considered $\gamma_{\rm h}=3\times 10^6$ (2.8 PeV), $10^7$ (9.4 PeV), $3\times 10^7$ (28 PeV), $10^8$ (94 PeV), and $3\times 10^8$ (280 PeV). We assumed $h=0.7 a$, $\theta_{\rm j}=0$, $i=28\degr$, and the stellar model from Section \ref{photon}. For these parameters, all photopion production occurs below the photospheric radius ($R<R_{2/3}$), and the minimum approach to the star is at $\approx\! 1.4 R_*$. The total interaction probabilities, given by Equation (\ref{prob}), are shown for the jet and counterjet in Figure \ref{tau_plot}(a). As expected, we find that the counterjet modulation depth is much higher than that for the jet. In Figure \ref{tau_plot}(b), we see that an approximate effective threshold for photopion production is ${\cal E}\sim 10$ PeV (corresponding to $\gamma_{\rm h}= 10^7$). 

The relative probabilities of interactions with the wind are shown in Figure \ref{tau_plot}(c). We assumed $\sigma_{\rm pHe}=200$ mb (independent of $\gamma_{\rm h}$). The weak dependence on $\gamma_{\rm h}$ follows from the definition of this probability as a channel of the total hadronic rate, which also depends on $\tau_{\pi,\gamma}$ (see Equation (\ref{probHe})). The relatively high probability of wind interactions follows from the adopted model's loss rate, $\dot M_{\rm w}=10^{-4.74}\msun$/yr. On average, 0.27 of the accelerated hadrons produce pions, and approximately 1/3 of the produced pions are neutral. However, the actual value of $\dot M_{\rm w}$ in Cyg X-3 remains uncertain, with the current best estimates around $\dot M_{\rm w}\approx 10^{-5}\msun$/yr (see \citet{ZMB13}, \citet{Koljonen17}, and references therein). In that case, the probabilities of this process would be about half of those presented here.

We then average the $\pi^0$ interaction probability, given by Equation (\ref{probpi0}), over a power-law proton distribution,
\begin{equation}
\langle{\cal P}_{\pi^0,\gamma}\rangle(p,\gamma_{\rm min},\gamma_{\rm max})=\frac{1-p}{\gamma_{\rm min}^{1-p} - \gamma_{\rm max}^{1-p}}
\int_{\gamma_{\rm min}}^{\gamma_{\rm max}}\!\! {\rm d}\gamma_{\rm h} \gamma_{\rm h}^{-p} {\cal P}_{\pi^0,\gamma}(\gamma_{\rm h}).\label{average}
\end{equation}
(Analogously, we can average other probabilities.) Figure \ref{tau_plot}d shows an example of the $\pi^0$ production rate for $p=2.2$ (\lh) with $\gamma_{\rm min}=10^7$ and $\gamma_{\rm max}=10^8$, for which the computed rate is roughly proportional to the PeV photon production rate detected by LHAASO. We see that the counterjet probabilities are significantly higher than those of the jet. The phase-averaged probability for the counterjet is $\approx$0.10.

\section{Energetic constraints}
\label{energetic}

The observed energy flux, corrected for CMB absorption (see \lh), in the 0.06--4 PeV range is $\approx\! 5\times 10^{-13}$ erg cm$^{-2}$ s$^{-1}$. Assuming isotropy at 9 kpc, this corresponds to $\approx\! 5\times 10^{33}$ erg s$^{-1}$. The luminosity for the 1--4 PeV range alone is $\approx\! 3.7\times 10^{33}$ erg s$^{-1}$, and the excess above the extrapolation of the power-law-like spectrum seen below 1 PeV to the 1--4 PeV range is $\sim\! 3 \times 10^{33}$ erg s$^{-1}$. We attribute this excess to $\pi^0$ photopion production. The average interaction probability for hadrons in the energy range ${\cal E}\approx 10$--100 PeV is 0.1 (Section \ref{depths}), and the inelasticity is $\sim$0.2 (Section \ref{hadron}). Thus, the power in that range, $P_{10-100}$, required to account for the observed emission is $(0.1\times 0.2)^{-1}$ times the photopion luminosity, i.e., $\sim\! 1.5\times 10^{35}$ erg s$^{-1}$, which is very modest compared to, e.g., the estimate of the magnetic power in Equation (\ref{PB}). The total power in relativistic hadrons will strongly depend on their distribution below 10 PeV, which contributes only minimally to the observed photopion-produced spectrum. For a power-law distribution with $p\neq 2$, we have the scaling of the fractional power in hadrons of
\begin{equation}
    \frac{P_{\rm h}}{P_{10-100}}= \frac{({\cal E}_{\rm min}/1\,{\rm PeV})^{2-p}-100^{2-p}}{10^{2-p}-100^{2-p}}.
    \label{Pratio}
\end{equation}
\lh modeled the observed spectrum with $p=2.2$. For example, at ${\cal E}_{\rm min}=10$ GeV, this ratio is $\approx$41, yielding a power in accelerated hadrons of $P_{\rm h}\approx 6 \times 10^{36}$ erg s$^{-1}$. The power required to accelerate the hadrons is $P_{\rm h}$ divided by the acceleration efficiency.

The emission below $E=1$ PeV, as well as its extrapolation above 1 PeV, can be explained by pion production from the same hadrons interacting with He nuclei in the stellar wind. For our adopted wind $\dot M_{\rm w}$, this process is efficient, with $\approx$0.27 of the accelerated hadrons producing $\pi^0$  (Section \ref{depths}) and an inelasticity of $\approx$0.6 (Section \ref{hadron}).

\subsection{An illustrative TeV-PeV orbital light curve}

To obtain an illustrative orbital light curve in the $E > 0.1$ PeV band, we adopt a simplified approach in which the phase-dependent photon flux is proportional to the proton population, weighted by the (energy- and phase-dependent) probability of neutral-pion production. This is analogous to Equation~(\ref{average}) but without normalization by the proton distribution, i.e., we retain absolute proton weighting rather than computing a spectrum-averaged probability.
\begin{equation}
\begin{split}
F_{\gamma}(\phi) =\;& Q_1 \int_{\gamma_{\rm min,1}}^{\gamma_{\rm max,1}} {\rm d}\gamma \, \gamma^{-p} \, {\cal P}_{\pi^0,\gamma}(\gamma,\phi) \\
&+\, Q_2 \int_{\gamma_{\rm min,2}}^{\gamma_{\rm max,2}} {\rm d}\gamma \, \gamma^{-p} \, {\cal P}_{\pi^0,\rm He}(\gamma,\phi) \, ,
\end{split}
\label{photon_flux}
\end{equation}
where ${\cal P}_{\pi^0,\rm{He}}(\gamma,\phi)$ is the probability of producing neutral pions in $p$--He interactions, $Q_{1,2}$ are normalization constants, and $p=2.2$ is assumed.

We include both photo-hadronic ($p\gamma$) and hadron-nuclear ($p$--He) channels, represented by the first and second integrals in Equation (\ref{photon_flux}), respectively. For the $p\gamma$ process, which dominates emission above $\sim 1$ PeV, and assuming that most interactions occur near the $\Delta$-resonance, $\gamma$-rays in the $1$--$10$~PeV range are produced by protons with $\gamma_{\rm min,1} \sim 10^7$ and $\gamma_{\rm max,1} \sim 10^8$. For the $p$--He channel, dominant below $\sim 1$~PeV, although the total pion multiplicity is large ($\sim 50$--$100$), only the leading pions contribute to the light curve band. These carry a fraction $\sim 0.1$ of the proton energy, implying $E \approx 0.05\,{\cal E}_p$, such that $\gamma$-rays in the $0.1$--$1$~PeV range originate from protons with $\gamma_{\rm min,2} \sim 2\times10^6$ and $\gamma_{\rm max,2} \sim 2\times10^7$.

For the interaction probabilities, we use the $\pi^0$-production quantities defined earlier in the text. For the $p\gamma$ channel, this corresponds to ${\cal P}_{\pi^0,\gamma}(\gamma,\phi)$ (Equation~(\ref{probpi0})). For the $p$--He channel, the quantity ${\cal P}_{\pi,{\rm He}}(\gamma,\phi)$ (Equation~(\ref{probHe})) gives the total probability of pion production and must be converted to a neutral-pion probability. While the overall pion population contains $\sim 1/3$ neutral pions, this fraction does not apply to the highest-energy (leading) pions relevant here. At TeV--PeV energies, leading pion production in interactions of protons and neutrons with He nuclei is dominated by the projectile fragmentation region, where charge conservation and valence-quark structure suppress one of the charged pion species, while neutral pions remain unsuppressed. This leads to an approximately equal partition between neutral and charged leading pions, yielding an effective neutral-pion fraction $f_{\pi^0,{\rm lead}} \sim 0.5$. We therefore adopt ${\cal P}_{\pi^0,\rm{He}}(\gamma,\phi) = 0.5\,{\cal P}_{\pi,{\rm He}}(\gamma,\phi)$. For this calculation, these interaction probabilities are evaluated for the counterjet, which dominates the emission.

The total light curve is taken as the sum of both components. Since the $p\gamma$ channel produces approximately one neutral pion per interaction, and the definition of ${\cal P}_{\pi^0,\rm{He}}$ effectively corresponds to the production of one leading neutral pion per interaction contributing to the light curve band, the resulting photon yields per interacting proton are comparable. We therefore adopt equal normalization factors, $Q_1 = Q_2$, and neglect any residual differences (e.g., due to propagation effects or detailed multiplicity) at this level of approximation. In practice, internal $\gamma\gamma$ absorption on stellar photons (important below $\sim 0.5$~PeV) and attenuation on the CMB (relevant above $\sim 0.3$~PeV) will affect the two components differently; these effects are not included here and are deferred to future detailed modeling.

\begin{figure}
\includegraphics[width=\columnwidth]{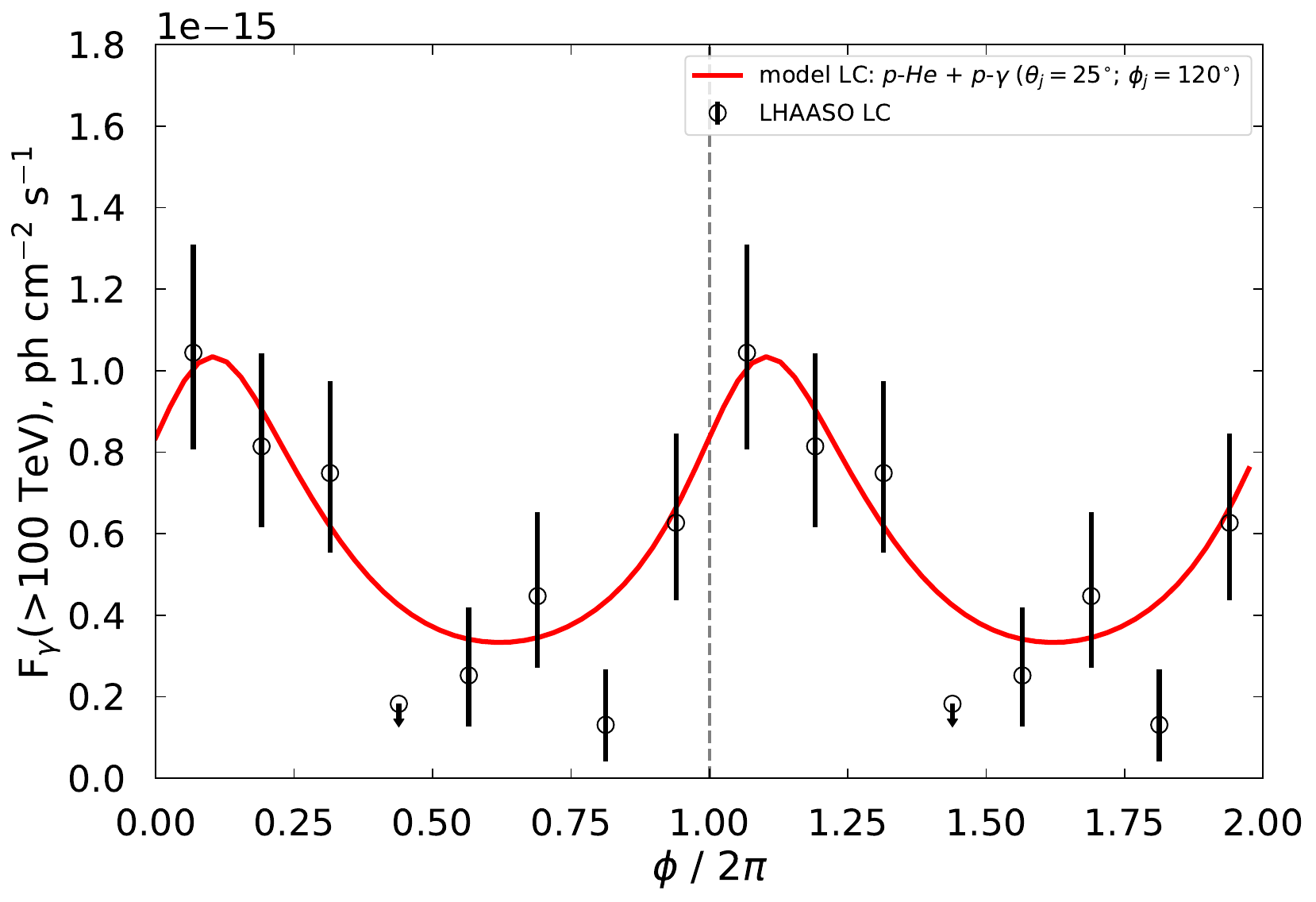}
\caption{An illustrative orbital light curve in the $E > 0.1$ PeV band (red solid line) is compared with the LHAASO data points (\lh). The model curve is computed using Equation (\ref{photon_flux}) with a proton spectrum having index $p=2.2$ and $Q_{1,2}=1.7\times10^{-7}$ cm$^{-2}$ s$^{-1}$. Both photo-hadronic and hadronic channels are included, but only the counterjet contribution is shown because it dominates the emission in this scenario. A misaligned jet geometry with $\theta_{\rm j}=25^\circ$ and $\phi_{\rm j}=120^\circ$ is assumed, with all other parameters identical to those in Figure \ref{tau_plot}. The model is intended for illustration only and provides a qualitative reproduction of the observed modulation pattern and phase offset, rather than a formal fit.
}\label{example_lc}
\end{figure}

For a jet aligned with the orbital axis ($\theta_{\rm j}=0$, $\phi_{\rm j}=0$), the emission is expected to peak at superior conjunction, in contrast to the phase offset observed by LHAASO. To reproduce this behavior, we consider a misaligned jet and find that $\theta_{\rm j}=25^\circ$ and $\phi_{\rm j}=120^\circ$ provide a good qualitative match, with the resulting example light curve shown in Figure \ref{example_lc}. In this configuration, with the counterjet dominating the emission, the geometric variation in the line-of-sight column density naturally accounts for both the modulation amplitude and the phase shift relative to superior conjunction.

\section{Neutrino flux estimate}
\label{neutrinos}

We estimate the all-flavor neutrino flux from the photo-hadronic ($p\gamma$) process in the energy range $E_\nu \sim 1$ PeV, where this process dominates. In the $\Delta$-resonance regime, each interaction transfers a fraction $K_{\pi}\approx 0.2$ of the proton energy to the pion. For charged pions, approximately $3/4$ of this energy is carried by neutrinos, yielding an effective fraction $K_\nu \approx 0.15$. With three neutrinos produced per interaction, the characteristic neutrino energy is $E_\nu \approx 0.05\,{\cal E}_{\rm p}$. For neutral pions, the decay $\pi^0 \to 2\gamma$ yields $E \approx 0.1\,{\cal E}_{\rm p}$, i.e., $E \approx 2 E_\nu$.

For the $\Delta$-resonance, the branching ratios are $f_{\pi^0}=2/3$ and $f_{\pi^+}=1/3$. Each interaction produces, on average, $2f_{\pi^0}=4/3$ photons and $3f_{\pi^+}=1$ neutrino. Thus, the neutrino number flux is $3/4$ of the $\gamma$-ray flux, and since $E \approx 2 E_\nu$ (for the same ${\cal E}_{\rm p}$), the corresponding relation between the neutrino and $\gamma$-ray SEDs is $E_\nu^2 \, {\rm d}N_\nu(E_\nu)/{\rm d}E_\nu \approx (3/8)  E^2 \, {\rm d} N_\gamma (E=2E_\nu)/ {\rm d} E$.

Using the (CMB-deabsorbed) LHAASO SED flux at $E \approx 2$ PeV of $\approx (3$--$4)\times 10^{-13}$ erg cm$^{-2}$ s$^{-1}$ (\lh), we obtain an all-flavor neutrino flux at $E_\nu \sim 1$ PeV of $F_\nu \approx (1.1$--$1.5)\times 10^{-13}$ erg cm$^{-2}$ s$^{-1}$. We neglect internal $\gamma\gamma$ absorption on stellar photons, which is expected to be subdominant above $\sim 1$ PeV. We emphasize that this estimate assumes that protons efficiently convert their energy into secondary particles, with interactions dominated by the $\Delta$-resonance, and it does not explicitly account for the energy- and phase-dependent interaction probabilities relevant in the optically thin regime.

Our PeV-scale estimate can be compared with constraints on the neutrino emission level from Cyg X-3 derived from IceCube data. Interpreting the two highest-significance neutrino candidate events reported by \citet{Koljonen23} as signal and folding them with the IceCube effective area yields an indicative neutrino flux of $\approx 5\times 10^{-12}$ TeV$^{-1}$ cm$^{-2}$ s$^{-1}$ at $\sim 1$ TeV (see also \citealt{Abbasi22} for a flux upper limit). Although a direct comparison is not straightforward because of the large energy separation and possible deviations from a simple power-law spectrum, our PeV-scale estimate remains broadly consistent with a neutrino flux that does not exceed the existing IceCube constraints at lower energies.

\section{Discussion and summary}
\label{discussion}

We have shown that the hadron acceleration region in the jet of Cyg X-3, which must be confined by strong magnetic fields, is spatially distinct from a much larger emission region with weaker magnetic fields. This conclusion follows from our measurements of the optical depths for pion production by accelerated hadrons on both stellar photons and the He nuclei of the WR donor's stellar wind, which we find to be $<1$, i.e., in the slow-cooling regime. This allows the accelerated hadrons to advect downstream in the jet to regions with weak magnetic fields, leave the jet, and then leave the binary.  

We have approximated the motion of accelerated hadrons as straight-line motion. This is a conservative assumption; accounting for the curvature of charged hadron trajectories along field lines would increase the pion production rate. Accounting for this effect would require knowledge of the magnetic field strength and structure. By contrast, accelerated He nuclei undergo photodissociation before producing pions (due to the larger photodissociation cross section and lower photodissociation threshold). Photodissociation produces neutrons, which travel in straight lines. 

Given the slow-cooling regime, there is no upper limit on the magnetic field strength in the acceleration region due to e$^\pm$ pair production by PeV photons interacting with the magnetic field, because very few such photons are produced there. 

In the slow-cooling regime, pion-production rates are proportional to the column densities along the particle path. The viewing angle of Cyg X-3 is very low, $\approx\!26\degr$--$28\degr$, as recently measured from X-ray polarization \citep{Veledina24b}. Consequently, the line-of-sight paths are much longer for the counterjet than for the jet. Therefore, most PeV $\gamma$-rays are produced by hadrons accelerated in the counterjet, provided the jet bulk motion is at most mildly relativistic (as measured, e.g., by \citealt{Dmytriiev24}).

In contrast, accelerated electrons lose energy rapidly when they scatter stellar photons, producing GeV photons; they are in the fast-cooling regime. Therefore, scattering occurs mostly in the acceleration regime. Given the boosting due to the jet's bulk motion, the jet emission dominates \citep{Dmytriiev24}. This can explain why the peak phase of the GeV-photon orbital modulation occurs on the opposite side of superior conjunction from that of the PeV photons. On the other hand, the GeV and PeV sources must be spatially separated, as follows from the upper limit on the magnetic field strength in the former being much lower than the lower limit on $B$ in the latter (Section \ref{regions}), as pointed out by \citet{Zhang26}.

In our calculations of the spatially dependent photopion production rate, we have taken into account details of the photon field around the WR donor star's hydrostatic core. The wind is optically thick, which causes a diffusive enhancement of its temperature below the photospheric radius. This results in a relatively complex radial temperature dependence. We stress that the commonly used parameters $R_*$ and $T_*$ correspond to the effective temperature of the hydrostatic core, i.e., the temperature the star would have in the absence of the wind. The actual radiation temperatures outside and inside the optically thick wind are substantially lower and higher than $T_*$, respectively, and we have taken this into account. Furthermore, we note an uncertainty in the actual values of $T_*$ and $R_*$, with different combinations of them producing the same observed spectrum \citep{Koljonen17}. 

Another major caveat affecting the accuracy of our formalism concerns the structure of the stellar photon field and the wind. We used a model of an isolated WR star. In reality, the stellar wind will be gravitationally focused toward the compact object and ionized by the X-ray source. As shown in the WR model we used, the wind of Cyg X-3 strongly overflows the Roche lobe of its WR star, whereas the recent model of \citet{White26} postulates even standard Roche lobe overflow through the L1 point. These modifications will increase the wind density in the equatorial plane, thereby increasing the rate of pion production by relativistic hadrons on He nuclei in the wind.

Also, we approximated the treatment of pion production in collisions of stellar photons by assuming that the photons follow radial trajectories. While the trajectories of photons above the photosphere can be approximated by radial paths, the distribution of photons below $R_{2/3}$ is closer to isotropic. Still, we do not expect this to introduce a major bias into the result, since the assumed photon trajectory can be considered an average one. We have also neglected the effect of e$^\pm$ pair absorption of the PeV photons on the stellar field (see \lh). 

Although we neglected jet misalignment in our study, it is included in our formalism, and we provided an example light curve that accounts for it. In our follow-up analysis, we will constrain the model parameter space using both the observed spectrum and the folded light curve, and we will account for pair absorption and photon trajectories. We will also refine our estimate of the expected neutrino flux.

Finally, we stress that we found the optical depths for hadronic pion production on both the stellar photon field and the stellar wind to be comparable to or less than unity. Since Cyg X-3 is by far the most compact known high-mass X-ray binary, these conditions point to the uniqueness of Cyg X-3 as a source of detectable PeV emission generated within the binary. For example, the binary separation of the high-mass X-ray binary Cyg X-1 is an order of magnitude larger than that of Cyg X-3, implying that the stellar emission seen from the jets is lower by two orders of magnitude. Furthermore, the mass loss rate of Cyg X-1 is significantly lower (e.g., \citealt{Gies03}). In low-mass X-ray binaries, both the stellar photon emission and the wind are much weaker than in Cyg X-3. This is consistent with the lack of unambiguous evidence that the PeV emission detected by LHAASO from other microquasars is associated with the immediate vicinity of the compact object \citep{LHAASO25}.

\section*{Acknowledgements}
We thank Cong Li and Krzysztof Nalewajko for valuable discussions and the referee for helpful comments. We acknowledge support from the Polish National Science Center under grants 2019/35/B/ST9/03944 and 2023/48/Q/ST9/00138. AD acknowledges support from the Department of Science, Technology, and Innovation and the National Research Foundation of South Africa through the South African Gamma-Ray Astronomy Programme (SA-GAMMA). KIIK has received funding from the European Research Council (ERC) under the European Union's Horizon 2020 Research and Innovation Programme (grant agreement No. 101002352, PI: M. Linares).

\bibliographystyle{aasjournal}
\bibliography{allbib.bib} 

\end{document}